\documentclass[journal,final,a4paper,twoside]{IEEEtran}
\usepackage{cite}

\ifCLASSINFOpdf
  \usepackage[pdftex]{graphicx}
  \graphicspath{{.}}
  \DeclareGraphicsExtensions{.pdf,.jpeg,.png}
\else
  \usepackage[dvips]{graphicx}
  \graphicspath{{.}}
  \DeclareGraphicsExtensions{.eps,.pdf}
\fi
\usepackage[cmex10]{amsmath}
\usepackage{array}

\usepackage{mdwmath}
\usepackage{mdwtab}
\usepackage{url}

\begin{document}
%
\title{$(MC^2)^2$: A Generic Decision-Making Framework 
and its Application to Cloud Computing}
%
%
%

\author{Michael Menzel, FZI Forschungszentrum Informatik Karlsruhe, menzel@fzi.de\\
        Marten Sch\"onherr, Telekom Laboratories Berlin, marten.schoenherr@telekom.de\\
        Jens Nimis, FZI Forschungszentrum Informatik Karlsruhe, nimis@fzi.de\\
        Stefan Tai, KIT Karlsruhe Institute of Technology, stefan.tai@kit.edu
}

%
%

\markboth{Short Version, No Example and Appendix}%
{Full Version Published in Proceedings of International Conference on Cloud Computing and Virtualization, GSTF, May 2010.}
%



\maketitle

\begin{abstract}
Cloud computing is a disruptive technology, representing a new model for information technology (IT) solution engineering and management that promises to introduce significant cost savings and other benefits. The adoption of Cloud computing requires a detailed comparison of infrastructure alternatives, taking a number of aspects into careful consideration. Existing methods of evaluation, however, limit decision making to the relative costs of cloud computing, but do not take a broader range of criteria into account. In this paper, we introduce a generic, multi-criteria-based decision framework and an application for Cloud Computing, the Multi-Criteria Comparison Method for Cloud Computing ($(MC^2)^2$). The framework and method allow organizations to determine what infrastructure best suits their needs by evaluating and ranking infrastructure alternatives using multiple criteria. Therefore, $(MC^2)^2$ offers a way to differentiate infrastructures not only by costs, but also in terms of benefits, opportunities and risks. $(MC^2)^2$ can be adapted to facilitate a wide array of decision-making scenarios within the domain of information technology infrastructures, depending on the criteria selected to support the framework.
\end{abstract}


%
\IEEEpeerreviewmaketitle

\section{Introduction}
%
%
%
%
\IEEEPARstart{T}{he} use of cloud computing services promises to eliminate large upfront capital investments and to reduce operational costs. Associated with these objectives are further benefits of improved innovation processes and a faster time-to-market. To which extent such cost savings and benefits can be achieved, however, and whether these are in conflict with other criteria, must be carefully evaluated. Criteria include technical criteria such as workload management patterns, scalability requirements, data volumes and data access patterns, as well as non-technical criteria ranging from strategic outsourcing models to regulatory compliance and legal issues. Alternatives to a Cloud-based solution are, for example, traditional on-premise IT infrastructures.

The decision whether or not to use Cloud computing services must take the diversity of relevant criteria into account. The set of criteria and their respective weight nevertheless varies from organization to organization, and from case to case. Therefore, a generic framework that can be tailored to specific organizations and application cases is needed.  The $(MC^2)^2$ method and framework introduced in this paper aims to address this problem and to fill the current void for a multi-criteria decision framework for evaluation and comparison of Cloud-based and non-Cloud IT infrastructure solutions.
In this paper, we introduce $(MC^2)^2$: a generic decision-making framework and its application for Cloud Computing, $(MC^2)^2$ fills the current void for a multi-criteria decision framework that supports a systematic evaluation and comparison of Cloud-based with non-Cloud IT infrastructure solutions.

The paper is structured as follows. After a reflection of related works in Section \ref{relatedWork}, we introduce the decision-making framework and its central process in Section \ref{mc22framework}. The framework describes an IT supported process that helps to build a method to evaluate IT solutions for specific scenarios. Next, we give an example of how to use the framework to define an evaluation method to compare different IT infrastructure options for the scenario of new software deployment. This example serves as a concrete use case that compares Cloud-based with non-Cloud, traditional IT infrastructure solutions. We discuss future work in Section \ref{discussfuture} and conclude in Section \ref{conclude}. 


 

\section{Related Work}
\label{relatedWork}
A common, recommended approach to evaluate arbitrary alternatives is to use multi-criteria decision methods. However, generic decision-making methods have to be customized for the domain of IT infrastructure solutions. Customization is non-trivial, as multi-criteria decision-making methods do not advise on how to find eligible criteria and factors that have an influence on the evaluation and decision making.
Current approaches to IT infrastructure decision-making that involve important criteria have been developed in operations research, in particular in the field of outsourcing theories \cite{lacity1993information}. Ngwenyama and Bryson \cite{ngwenyama1999making} introduced a transaction costs-based approach that allows to analyze information systems outsourcing decisions. However, this approach only supports a limited set of relevant factors, and is not open to extensions needed  to incorporate new criteria for outsourcing alternatives in the field of cloud computing. There have also been attempts to evaluate IT infrastructures and especially to support the decision making of whether to move information systems into the cloud or not. Two different approaches are described by Armbrust et al. \cite{armbrust2009above} and Walker \cite{10.1109/MC.2009.135}. Armbrust et al. propose a simple formula to compare the costs of a cloud service with those of a datacenter based on hours of use, whereas Walker proposes a net present value based approach to compare the costs of leasing and purchasing a CPU over several years. Both works focus only on a single cost aspect in the broad spectrum of relevant IT infrastructure criteria.

In our previous work, we proposed the CloudTCO framework \cite{klems2008dccaffetvocc} to widen the range of costs considered when comparing IT infrastructure options. CloudTCO, however, is limited to addressing non-technical requirements and focuses on costs, but does not compare benefits. Moreover, the resulting values can hardly be compared in a ratio scale but more in an interval or ordinal scale. Thus, concluding comparison statements such as "alternative 1 is two times better that alternative 2" are impossible to express.


\section{The $(MC^2)^2$ Framework}
\label{mc22framework}
The $(MC^2)^2$ framework is an evolution of the CloudTCO approach introduced by Klems et al. \cite{klems2008dccaffetvocc}. $(MC^2)^2$ extends the basic CloudTCO idea by supporting a wider range of diverse criteria, including benefitial factors . Further, while CloudTCO uses an utility function to compare result values $(MC^2)^2$ allows to evaluate IT infrastructure alternatives on a ratio scale. $(MC^2)^2$ has been designed as a generic framework that can be customized to define different, evaluation methods that are focused in scope and address a diverse set of criteria.

$(MC^2)^2$ proposes an abstract, linear process to step-wise define a concrete evaluation method. The process incorporates several steps from scenario and alternatives definition to evaluation and ranking. Additionally, the process includes the consideration of an organization�s internal and external information resources, and considers the integration and use of various software systems that can support the process, for example: 
\begin{itemize}
\item Knowledge bases within the organization, e.g., wikis, reporting and document management software systems, databases with explicit knowledge
\item Business intelligence systems, e.g., data mining and text mining tools
\item Survey creation, deployment and analysis software systems
\end{itemize}
The process suggests a step-by-step order to build a concrete, customized evaluation method. First, a scenario and a number of alternatives are defined. Next, relevant criteria and requirements are identified. A multi-criteria decision-making method is then chosen and subsequently configured; this sets the criteria and requirements as parameters of the selected decision-making method. The result is a custom evaluation method that can be used to evaluate the alternatives under consideration.

\begin{figure}[!t]
\centering
\includegraphics[width=2.5in]{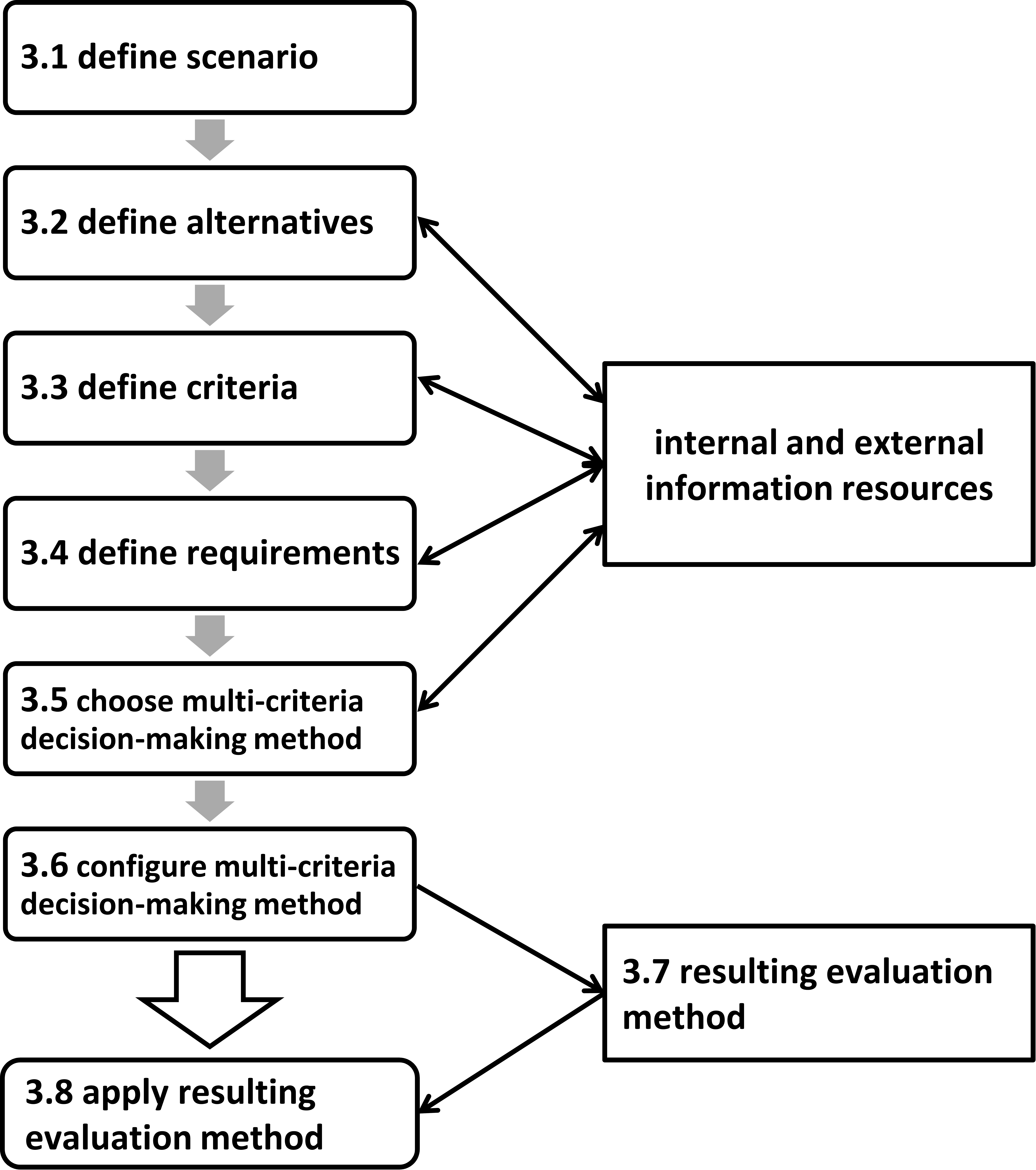}
\caption{The Process within the $(MC^2)^2$ Framework}
\label{fig1}
\end{figure}

Figure \ref{fig1} depicts the process steps, each of which is described in further detail below.  

\subsection{Define Scenario}
A scenario specifies the particular situation under consideration, describing the business and IT context, environment, constraints and requirements, and at the same time setting the goals. The description of the scenario sets the scope and goals for the evaluation method to be created. Comprehensive expertise and experience is required to define the scenario, and sufficient effort should be spent to facilitate the subsequent method steps.

\subsection{Define Alternatives}
When using our framework, at least two alternatives must be defined for a given scenario. Every alternative aims at being a solution to the scenario�s goals. Each alternative is defined by many attributes. An attribute is a name of a characteristic common to all alternatives and can be assigned a value. Every pair of alternatives differs in at least one attribute. Otherwise, an additional attribute in which both alternatives differ must be introduced. 

To derive alternatives the given situation and defined goals must first be studied. Finding alternatives for a scenario involves expertise, creativity and experience, which ideally is pooled organization-wide. This can be supported by IT systems such as:
\begin{itemize}
\item Expert search systems to set up expert panels or for consulting purposes
\item Work collaboration systems to collaboratively find alternatives
\item Internal databases with externalized knowledge about experiences, for instance:
	\begin{itemize}
	\item Pages in an internal wiki system about related topics
	\item Internal documents with past experiences and reports about related decisions and evaluations
	\end{itemize}
\end{itemize}
It is possible to perform this search process systematically by first searching for a set of important attributes and then assigning values to each attribute, for each alternative. At the end, each alternative is defined by the values assigned to the set of attributes. A value for an attribute can be an exact measured value, or an explanation or description in natural language. 
A detailed scenario description can significantly ease the search for attributes. A comprehensive scenario description may already reveal different options and variables within the scenario. Each option or variable is a potential attribute. 

Besides analyzing the scenario�s description, again expertise and experience usually leads to the identification of many important attributes. By considering information resources from within or from outside the organization the attribute identification process can make use of existing externalized knowledge. (For reasons of focus, this paper, however, will not examine possible information resources and specific knowledge management techniques useful to externalize experiences made during the process.)

\subsection{Define Criteria}
Alternatives defined by attributes do not necessarily provide sufficient information for an evaluation. The introduction of criteria extends attribute definitions with scales and makes them usable in evaluation methods. This is why one of the essential steps within $(MC^2)^2$ is the definition of criteria. Having completed this step, requirements can be derived from the criteria and the multi-criteria decision-making method of choice can be configured and applied.

The set of attributes determined during the definition of alternatives serve as a starting point to derive the set of criteria. Criteria are different from attributes in their form. A criterion consists of a topic or question to be examined and a type which is either qualitative or quantitative. 
In case of a quantitative criterion a scale of measurement has to be defined. Expect some multi-criteria decision-making methods to restrain the choice of measurement scales. This implies that depending on the multi-criteria decision-making method of choice quantitative criteria that do not fulfill the scale restrictions will have to be excluded from the set of criteria and, thus, will not be considered by the resulting evaluation method. 

Qualitative criteria do not necessarily rely on a scale as they are not easily measurable. Some multi-criteria decision-making methods do not support qualitative criteria and require transforming qualitative criteria into quantitative criteria by creating a scale.

Furthermore, criteria do not always influence a decision positively but negatively as well. According to Saaty \cite{saatybook2005} criteria can typically be clustered into the four types benefits, opportunities, costs, and risks. While criteria of the types benefits or opportunities are positive, criteria of the types costs and risks are negative. Nevertheless, other criteria type categories are possible as well.
Table 1 shows four example criteria where criterion \#1 is a quantitative criterion with nominal scale, \#2 a quantitative criterion with binary nominal scale, \#3 is a quantitative criterion with ratio scale, and \#4 is a qualitative criterion.
\begin{table}[!t]
\renewcommand{\arraystretch}{1.3}
\caption{Example Criteria}
\label{table_examplecriteria}
\centering
\begin{tabular}{| c | m{1in} | m{.7in} | m{.5in} |}
\firsthline
\# & Question & Type & Possible Values\\
\hline
1 & How high is the data security level of the physical storage? & quantitative, nominal scale & low, medium, high \\
\hline
2 & Is the usage of a SOA possible within the IT infrastructure? & quantitative, binary nominal scale & yes, no \\
\hline
3 & What are the monthly costs for upstream network traffic from the infrastructure location to the organization? & quantitative, ratio scale & 0\$-$\infty$\$ \\
\hline
4 & How friendly and helpful is the support given by the infrastructure provider? & qualitative & --- \\
\lasthline
\end{tabular}
\end{table}

To strive to an optimal evaluation of all alternatives an extensive set of criteria is favorable and facilitates more precise evaluation results. Pardee and Kirkwood \cite{pardee1969measurement} give three objectives to be pursued during the finding of criteria.
\begin{enumerate}
\item Completeness and Exhaustiveness
\item Mutually exclusive items only
\item Restrict to criteria of highest degree of importance
\end{enumerate}

Again, expertise and creativity are helpful to identify eligible criteria. Keeney and Raiffa \cite{keeney1993decisions} suggest research in literature and to consult expert panels in order to find further criteria. Moreover, finding and defining criteria can be supported by IT-based information resources and software systems. Gathering experts into panels can be supported by expert search systems. Information resources to derive criteria from include:
\begin{itemize}
\item Guidelines from departments or externalized tacit knowledge inside the company, for instance, guidelines of the controlling department about cost hierarchies for IT infrastructures
\item Results from surveying experts of different fields within the organization such as IT experts, security experts, controlling experts, or finance experts
\end{itemize}

\subsection{Define Requirements}
Prior to the evaluation step, requirements are defined. Requirements help to filter out alternatives that are not feasible within the constraints of the given scenario. A requirement of a scenario is expressed as a minimal or maximal constraint that has to be attained by an alternative regarding one criterion. Every alternative with at least one criterion value lower than a minimum or higher than a maximum constraint is not realizable and, hence, filtered out. An alternative that has been filtered out will not be considered in the evaluation method.

The procedure of filtering can be performed for minimum constraints with a conjunctive satisficing method and respectively for maximum constraints with a disjunctive satisficing method \cite{yoon1995multiple}. Typically, the conjunctive method is applied to restrictions on positive criteria and the disjunctive method to restrictions on negative criteria.
Optionally, there might be restrictions that should not be considered as criteria, but as a requirement only. In this case a new criterion needs to be introduced which is only considered as a requirement, but not as a criterion in the set of criteria.

As in criteria definition IT-based information resources and software systems are capable of supporting this step.

\subsection{Choose Multi-Criteria Decision-Making Method}
Many multi-criteria decision-making methods are eligible upon which the custom evaluation method can be defined. An overview and comparison of different multi-criteria decision-making methods can be found in \cite{baker2001guidebook}\cite{yoon1995multiple}. The method has to be chosen according to the preferences of an organization and depending on the case. It is required that application of the method results in values, and not just a ranking. 

We suggest the Analytic Network Process (ANP) \cite{saatybook2005}\cite{saaty2004decision} method as the favorable evaluation method due to its ability to incorporate complex criteria networks during the evaluation. More complex criteria networks enable a more realistic modeling of criteria dependencies. In ANP, criteria networks not only include weights, but feedback loops and clusters as well. Moreover, ANP employs pair-wise comparisons and normalization to support the use of qualitative criteria. The combination of pair-wise comparisons and normalization succeeds to assign values to quantitative and qualitative criteria on a ratio scale. However, qualitative criteria values still are derived from subjective ratings.

\subsection{Configure Multi-Criteria Decision-Making Method}
\label{configureMCDM}
Before an evaluation method can be build and the evaluation results can be computed with that method in the final step of the process, it may be required to configure the multi-criteria decision-making method. This typically means setting parameters such as alternatives, requirements, and criteria, and defining weights and relations for the set of criteria. Whether a configuration is needed or not depends on the evaluation method of choice. 

Setting parameters for an evaluation method mostly means balancing parameters to customize the method to fit the needs of the scenario and goals. For most multi-criteria decision-making methods this step consists of defining criteria weights and relations. In doing so, many multi-criteria decision-making methods allow setting focus on several criteria within the set of criteria by assigning weights. Thereby the influence of a criterion is intensified and other criteria's influence is weakened. The weight describes the relative importance of a criterion compared to all other criteria.

The process of defining criteria interdependencies, and in particular criteria networks can also be supported by IT based information resources and software systems, especially when complex criteria networks are desired to model the real world situation. Moreover, repositories of criteria network patterns are helpful.

\subsection{Resulting Evaluation Method}
\begin{figure}[!t]
\centering
\includegraphics[width=2.5in]{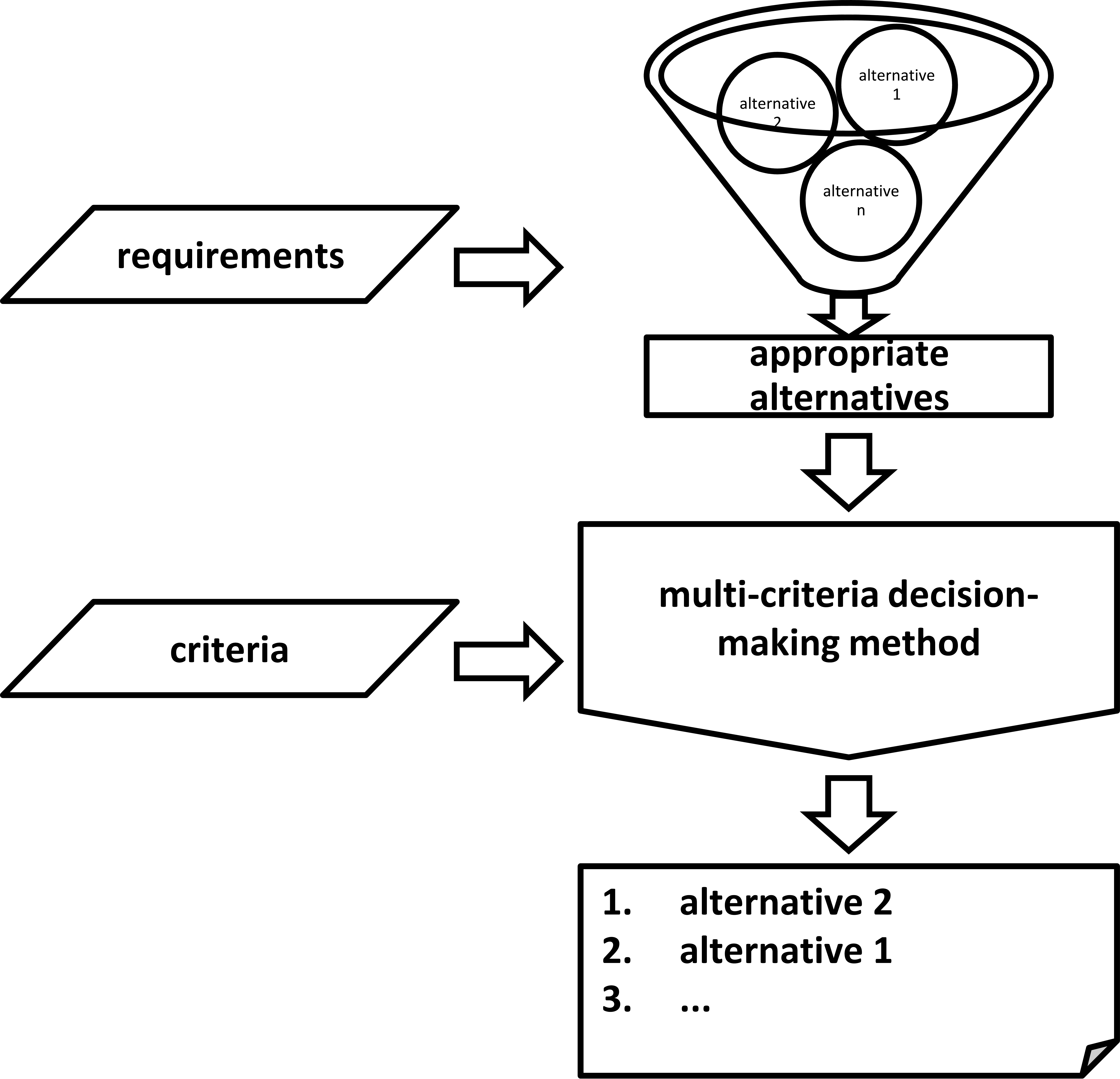}
\caption{Process of a Resulting Evaluation Method }
\label{fig2}
\end{figure}

The process of the resulting rational evaluation method and its inputs is depicted in Figure \ref{fig2}: Requirements are used to filter out all alternatives that cannot fulfill each of the requirements. Following, all remaining alternatives are evaluated by the given criteria considered in the multi-criteria decision-making method of choice. To rank the alternatives the alternatives can be sorted by their evaluation results. The alternative ranked as \#1 is the optimal decision according to its overall value. 

The resulting evaluation method is reusable with changing alternatives for decisions in the defined scenario's context.

\subsection{Apply Resulting Evaluation Method}
In the final step the resulting, custom evaluation method is applied to filter out the realizable alternatives and calculate the evaluation results. The calculation results in a rank and a value assignment for each of the alternatives. The assigned value is according to the method of choice either absolute or a ratio value relative to the other alternatives. We suggest to choose a multi-criteria decision-making method that returns values on a ratio scale.

The final result of the custom evaluation method can be used to make a substantiated, rational decision for an alternative representing a solution in a scenario.
Section 8 in the appendix examines the Analytic Network Process (ANP) which is a comprehensive evaluation method that provides a ranking of alternatives and value assignments on a ratio scale when applied within a custom method created with the $(MC^2)^2$ framework. The ANP serves not only as an example for a possible multi-criteria decision-making method that can be applied within a custom method�s process, but is also the favorable multi-criteria decision-making method that we suggest. As mentioned in Subsection \ref{configureMCDM} ANP allows to define complex criteria networks and it results in an evaluation on a ratio scale, which makes it a favorable basis for an evaluation method.

\section{Discussion \& Future Work}
\label{discussfuture}
The use of ANP to evaluate alternatives of IT infrastructures raises several questions. First of all, pair-wise comparisons with qualitative criteria are not trivial and must be done subjectively as qualitative criteria are not measurable. Group evaluations might be a successful approach to strive for more objective results in evaluations with many qualitative criteria. This is still open for future work.

Also, values of measurable quantitative criteria must be collected first. Support for this step is not included in the framework yet. Later works to intensify the IT support within the framework might include a support for data measurement and collection or a list of adequate information resources might be provided. Nevertheless, further research on IT support for the framework is needed. In particular, we aim to design easy-to-use systems that support $(MC^2)^2$ and which integrates a set of relevant existing information systems as information resources or tools. Especially, a list of suitable information systems that serve as information resources to support the definition of alternatives, criteria and requirements is an important future task. This should also include a categorization of the information resources regarding different aspects, like e.g., their usefulness.

Furthermore, fields of future work include research of applicable sensitivity analysis tools and the integration into the $(MC^2)^2$ framework, and a comparison of the advantages and applicability of different multi-criteria decision-making methods. Moreover, an evaluation of the framework is an important future step.

\section{Conclusions}
\label{conclude}
Cloud Computing presents an opportunity to organizations of all kinds, the benefits, risks, and promises of which must be carefully considered. Adopting Cloud Computing is a decision-making problem that requires identification of criteria and value-driven comparison of alternatives with respect to the criteria selected.
In this paper, we introduced $(MC^2)^2$: a generic decision-making framework and an evaluation method for Cloud Computing as its instantiation. $(MC^2)^2$ is a multi-criteria decision framework that defines a systematic, step-wise process that can be tailored to specific needs and preferences.

To our knowledge, $(MC^2)^2$ is the first comprehensive decision making framework and method specifically in support of Cloud Computing scenarios. Nevertheless, $(MC^2)^2$ is not restricted to Cloud Computing or even IT infrastructure solutions, but has been designed as a generic approach that can be applied to other complex decision-making processes also.


%





\ifCLASSOPTIONcaptionsoff
  \newpage
\fi



\bibliographystyle{IEEEtran}
\bibliography{IEEEabrv,referenzen}
%


%








\end{document}